\documentclass[trackchanges]{aastex701}

\usepackage{tabularx}  % in preamble
\usepackage{array}     % for better column control
\usepackage{makecell}
\usepackage{amsmath} % For math formatting, optional

\shortauthors{A. Gupta et al.}

\begin{document}

\title{Multi-directional investigations on quiet time suprathermal ions measured by ASPEX-STEPS on-board Aditya L1}

\author[orcid=0009-0000-9917-2694,sname='Gupta']{Aakash Gupta}
\altaffiliation{Physical Research Laboratory (PRL), Ahmedabad-380009, India}
\affiliation{Physical Research Laboratory (PRL), Ahmedabad-380009, India}
\affiliation{Indian Institute of Technology (IIT), Gandhinagar-382055, India}
\email[show]{aakashgupta.du@gmail.com}  

\author[orcid=0000-0003-2693-5325, sname='Chakrabarty']{Dibyendu Chakrabarty} 
\affiliation{Physical Research Laboratory (PRL), Ahmedabad-380009, India}
\email{dipu@prl.res.in}

\author[orcid=0000-0002-2050-0913, sname='Vadawale']{Santosh Vadawale}
\affiliation{Physical Research Laboratory (PRL), Ahmedabad-380009, India}
\email{santoshv@prl.res.in}

\author[orcid=0000-0002-4781-5798, sname='Sarkar']{Aveek Sarkar}
\affiliation{Physical Research Laboratory (PRL), Ahmedabad-380009, India}
\email{aveeks@prl.res.in}

\author[orcid=0000-0001-8993-9118, sname='Dalal']{Bijoy Dalal}
\affiliation{Physical Research Laboratory (PRL), Ahmedabad-380009, India}
\email{bijoydalal.at@gmail.com}

\author[orcid=0000-0002-3153-537X, sname='Goyal']{Shiv Kumar Goyal}
\affiliation{Physical Research Laboratory (PRL), Ahmedabad-380009, India}
\email{goyal@prl.res.in}

\author[orcid=0009-0004-5904-5833, sname='Sebastian']{Jacob Sebastian}
\affiliation{Physical Research Laboratory (PRL), Ahmedabad-380009, India}
\email{jacobs@prl.res.in}

\author[orcid=0000-0003-2504-2576, sname='Janardhan']{P. Janardhan}
\affiliation{Physical Research Laboratory (PRL), Ahmedabad-380009, India}
\email{Janardhan.Padmanabhan@gmail.com}

\author[orcid=0000-0002-0452-5838, sname='Srivastava']{Nandita Srivastava}
\affiliation{Udaipur Solar Observatory (USO), Physical Research Laboratory, Udaipur–313001, India}
\email{nandita@prl.res.in}

\author[orcid=0000-0002-5995-8681, sname='Shanmugam']{M. Shanmugam}
\affiliation{Physical Research Laboratory (PRL), Ahmedabad-380009, India}
\email{shansm@prl.res.in}

\author[orcid=0000-0003-4269-340X, sname='Tiwari']{Neeraj Kumar Tiwari}
\affiliation{Physical Research Laboratory (PRL), Ahmedabad-380009, India}
\email{neeraj@prl.res.in}

\author[orcid=0000-0001-8900-3635, sname='Sarda']{Aaditya Sarda}
\affiliation{Physical Research Laboratory (PRL), Ahmedabad-380009, India}
\email{aaditya@prl.res.in}

\author[orcid=0000-0001-9670-1511, sname='Sharma']{Piyush Sharma}
\affiliation{Physical Research Laboratory (PRL), Ahmedabad-380009, India}
\email{piyush@prl.res.in}

\author[orcid=0000-0003-1693-453X, sname='Bhardwaj']{Anil Bhardwaj}
\affiliation{Physical Research Laboratory (PRL), Ahmedabad-380009, India}
\email{bhardwaj_spl@yahoo.com}

\author[orcid=0000-0001-7590-1311, sname='Kumar']{Prashant Kumar}
\affiliation{Physical Research Laboratory (PRL), Ahmedabad-380009, India}
\email{prashantk@prl.res.in}

\author{Manan S. Shah}
\affiliation{Physical Research Laboratory (PRL), Ahmedabad-380009, India}
\email{manans@prl.res.in}

\author[orcid=0000-0002-7248-9859, sname='Bapat']{Bhas Bapat}
\affiliation{Indian Institute of Science Education and Research (IISER), Pune-411008, India}
\email{bhas.bapat@iiserpune.ac.in}

\author{Pranav R. Adhyaru}
\affiliation{Physical Research Laboratory (PRL), Ahmedabad-380009, India}
\email{pranav@prl.res.in}

\author[orcid=0000-0002-0929-1401, sname='Patel']{Arpit R. Patel}
\affiliation{Physical Research Laboratory (PRL), Ahmedabad-380009, India}
\email{arpitp@prl.res.in}

\author[orcid=0000-0002-5272-6386, sname='Adalja']{Hitesh Kumar Adalja}
\affiliation{Physical Research Laboratory (PRL), Ahmedabad-380009, India}
\email{hladalja@prl.res.in}

\author[orcid=0009-0000-9287-6154, sname='Kumar']{Abhishek Kumar}
\affiliation{Physical Research Laboratory (PRL), Ahmedabad-380009, India}
\email{abhishekkumar@prl.res.in}

\author[orcid=0000-0001-6022-8283, sname='Ladiya']{Tinkal Ladiya}
\affiliation{Physical Research Laboratory (PRL), Ahmedabad-380009, India}
\email{tinkal@prl.res.in}

\author[orcid=0009-0004-2604-9635, sname='Kumar']{Sushil Kumar}
\affiliation{Physical Research Laboratory (PRL), Ahmedabad-380009, India}
\email{sushil@prl.res.in}

\author[orcid=0000-0002-1975-0552, sname='Singh']{Nishant Singh}
\affiliation{Physical Research Laboratory (PRL), Ahmedabad-380009, India}
\email{nishant@prl.res.in}

\author[orcid=0009-0009-8350-491X, sname='Painkra']{Deepak Kumar Painkra}
\affiliation{Physical Research Laboratory (PRL), Ahmedabad-380009, India}
\email{deepakp@prl.res.in}

\author[orcid=0009-0009-7139-9112, sname='Verma']{Abhishek J. Verma}
\affiliation{Physical Research Laboratory (PRL), Ahmedabad-380009, India}
\email{abhishekv@prl.res.in}

\author{Swaroop Banerjee}
\affiliation{Physical Research Laboratory (PRL), Ahmedabad-380009, India}
\email{swaroopsb@gmail.com}

\author{K. P. Subramanian}
\affiliation{Physical Research Laboratory (PRL), Ahmedabad-380009, India}
\email{Aniyan.Kurur@gmail.com}

\author{M. B. Dadhania}
\affiliation{Physical Research Laboratory (PRL), Ahmedabad-380009, India}
\email{mbdprl@gmail.com}

\begin{abstract}
The origin, acceleration and anisotropy of suprathermal ions in the interplanetary medium during quiet time have remained poorly understood issues in solar wind physics. Using measurements (in the energy range of $0.12 - 1.33$ MeV/n) by the four detectors that are part of the Supra-Thermal and Energetic Particle Spectrometer (STEPS) of Aditya Solar Wind Particle EXperiment (ASPEX) on-board Aditya L1 spacecraft, we address the variations in spectral indices with directions in shorter durations during January–November 2024, which coincides with the maximum phase of Solar Cycle 25. Three out of four detectors on STEPS - Parker Spiral (PS), Inter-Mediate (IM), Earth Pointing (EP) - are in one plane while the fourth detector - North Pointing (NP) – is in a mutually orthogonal plane. The derived spectral indices are found to be $-1.99 \pm 0.06$ regardless of directions suggesting directionally near isotropic behavior during quiet times. It is also shown that the influence of Compton-Getting effect is negligible in this assessment of directional isotropy. This result has important ramification as directional isotropy is assumed while solving the Parker transport equation to explain the acceleration of energetic particles. Further analysis of elemental abundance ratios (${}^3\mathrm{He}/{}^4\mathrm{He}$, $\mathrm{Fe}/\mathrm{O}$, and $\mathrm{C}/\mathrm{O}$) during the same quiet times obtained from the Ultra-Low Energy Isotope Spectrometer (ULEIS) on –board the Advanced Composition Explorer (ACE) spacecraft suggests possible contributions from the leftover ions from the previous solar energetic particle (SEP) events in the quiet time suprathermal ion pool.

\end{abstract}

\keywords{\uat{Solar energetic particles}{1491} --- \uat{Solar wind}{1534} --- \uat{Solar coronal mass ejections}{310} --- \uat{Solar flares}{1496} ---\uat{Interstellar medium}{847}}

\section{Introduction} 

An important feature of the solar wind is the suprathermal ion tail which is manifested in the form of ions in the energy range of $\sim10~\mathrm{keV~n^{-1}}$ to $\sim1~\mathrm{MeV~n^{-1}}$. These suprathermal ions are ubiquitous in the heliosphere and are continuously observed to arrive at the spacecraft location from multi-directions in the interplanetary (IP) medium. The suprathermal ions act as a seed population for solar energetic particles (SEPs) accelerated by IP shocks associated with transient events in the heliosphere such as coronal mass ejections (CMEs; \citealt{Gosling_et_al_1981, Mason_et_al_1999, Desai_et_al_2003, Desai_et_al_2004, Chakrabarty_et_al_2025}) and corotating interaction regions (CIRs; \citealt{Fisk_and_Lee_1980, Chotoo_et_al_2000, Kucharek_et_al_2003, Mason_et_al_2008, Allen_et_al_2019}). In astrophysical plasmas, particle acceleration is primarily governed by two well-established mechanisms: first-order Fermi acceleration, also known as diffusive shock acceleration \citep{Krymskii_1977, Bell_1978}, and second-order Fermi acceleration, originally proposed by \citet{Fermi_1949}. Both mechanisms require the existence of a pre-accelerated ion seed populations whose energies are above the injection threshold and which can be further energized within the acceleration framework. Therefore, the importance of suprathermal ion pool is undeniable to understand the generation of SEPs.

The IP medium hosts a diverse reservoir of suprathermal ions, consisting of protons as well as heavier ions ($^4$He to Fe and beyond) \citep{Desai_and_Giacalone_2016}. Studies of the composition of these suprathermal ion populations indicate that their origins can be traced to multiple sources. These sources include solar wind  (e.g., \citealt{Desai_et_al_2003}), remnants from the past transient events such as CMEs and solar flares \citep{Mason_et_al_1999, Reames_2013}, and interstellar pick-up ions \citep{Allen_et_al_2019}. Notably, beyond 1 AU, interstellar pick-up ions have been found to contribute into the suprathermal ion pool \citep{Fisk_1976}. Their abundance and energy distribution play a crucial role in shaping the overall dynamics of suprathermal ion population and contribute to the seed population for the subsequent shock acceleration in the heliosphere \citep{Tylka_et_al_2001}. Suprathermal ions typically exhibit power-law spectra when the differential directional flux (phase space density) is plotted as a function of energy (speed). In general, a power law spectrum and its spectral index (slope of the power spectrum) correspond to a natural stochastic process. In case of suprathermal particles also, theoretical models such as pump acceleration mechanism suggests that this power-law spectrum follows a characteristic spectral index of approximately $-1.5\ (-5)$ in the differential directional flux vs. energy (phase space density vs. speed) representation \citep{Fisk_and_Gloeckler_2006, Fisk_and_Gloeckler_2008, Fisk_and_Gloeckler_2012, Fisk_and_Gloeckler_2014} during “quiet” times. However, several studies (e.g. \citealt{Dayeh_et_al_2009, Dayeh_et_al_2017, Dalal_et_al_2022}) show that the spectral index of suprathermal ions in the 100 keV/n to 1 MeV/n energy range during “quiet” times is not universally fixed at $-1.5$, but varies over a wide range ($- 1.2$ to $– 2.9$ ). Moreover, these works also reveal that, on many occasions, different elements have significantly different spectral indices during a given interval suggesting $m/q$ dependence as far as acceleration mechanisms are concerned (e.g. \citealt{Dalal_et_al_2022}).   Therefore, the understanding on the source and energization process(es) of suprathermal ions in the IP medium is far from complete. 

While dealing with acceleration of suprathermal ions in the interplanetary (IP) medium, Parker transport equation (PTE) is solved on many occasions with the assumption that suprathermal ions are isotropic in the IP medium (e.g. \citealt{Fisk_and_Gloeckler_2008}). Since suprathermal ion fluxes obtained from the Supra-Thermal and Energetic Particle Spectrometer (STEPS) of Aditya Solar Wind Particle EXperiment (ASPEX) on-board India’s Aditya L1 are directionally resolved, experimental verification of the isotropic assumption is possible based on evaluation of spectral indices. Further, unlike previous works (e.g. \citealt{Dayeh_et_al_2009, Dayeh_et_al_2017, Dalal_et_al_2022}) where datasets for longer intervals are used, we use shorter intervals of data from ASPEX-STEPS for this study to minimize influence of multiple transient processes in the suprathermal ion fluxes. After selecting shorter intervals (a few days) of quiet period fluxes in 2024, we check the variations in spectral indices with direction. 

It is to be noted here that although directionally resolved fluxes are obtained from ASPEX-STEPS, these fluxes are not species separated and are primarily dominated by protons and alpha particles. Therefore, in order to understand the source process(es) through which these quiet time suprathermal ions are generated, elemental abundance data for the same quiet time is analyzed. We use the data from Ultra Low Energy Isotope Spectrometer (ULEIS) on –board the Advanced Composition Explorer (ACE) spacecraft for this purpose.

Therefore, we intend to address two problems in this work. One, whether the measured suprathermal ion fluxes are directionally isotropic at L1 and two, what are the possible sources of these ions. The manuscript is organized as follows.  The datasets are discussed in section 2. In Section 3, the methodology used to select the quiet time is discussed. The results are presented in Section 4.  In Section 5, we discuss the results and in Section 6, the inferences from this work are summarized.  

\section{Dataset} \label{sec:Dataset}

Aditya L1 \citep{Seetha_and_Megala_2017, Tripathi_et_al_2022, Parate_et_al_2025} is the first observatory class mission of India to study the Sun and solar wind.  Aditya L1 was launched on 2nd September 2023 and placed in the halo orbit around the first Lagrange point (L1) of the Sun-Earth system on 6th January 2024. The Aditya Solar wind Particle EXperiment (ASPEX) \citep{Janardhan_et_al_2017, Goyal_et_al_2018, Jsebastian_et_al_2025} is one of three in-situ experiments that measures the solar wind  and energetic ions. ASPEX comprises two subsystems: the Solar Wind Ion Spectrometer (SWIS) \citep{Kumar_et_al_2025, Akumar_et_al_2025} and the Supra-Thermal and Energetic Particle Spectrometer (STEPS) \citep{Goyal_et_al_2025, sebastian_et_al_2025}. STEPS is designed to measure suprathermal and energetic ions. There are six detectors in STEPS – Sun-Radial (SR), Inter-Mediate (IM), Parker Spiral (PS), North Pointing (NP), South Pointing (SP) and Earth Pointing (EP).  Among the six detectors, the SR and SP detectors are experiencing light saturation issues and data from these detectors are not used in this work. As described in \citet{Goyal_et_al_2025, sebastian_et_al_2025}, the NP and IM units are a single-window detector equipped with a dead layer of $0.2\,\mu\mathrm{m}$ thickness. In contrast, the PS and EP units consist of a stack of custom-built dual-window Si-PIN (Silicon Positive-Intrinsic-Negative) detector and a combined scintillator plus silicon photomultiplier (SiPM) detector assembly. The dual-window Si-PIN detector features two active regions with different dead layer thicknesses: $0.1\,\mu\mathrm{m}$ for the circular inner detector (PS-Inn and EP-Inn, with a diameter of $7\,\mathrm{mm}$) and $0.8\,\mu\mathrm{m}$ for the annular outer detector (PS-Out and EP-Out, spanning diameters from $7\,\mathrm{mm}$ to $18\,\mathrm{mm}$). Due to the extremely thin dead layer in the PS-Inn and EP-Inn detector, it exhibits distinct spectral signatures corresponding to detection of multiple ion species. As the deconvolution of these spectral features in the PS-Inn and EP-Inn detector is still ongoing, this study utilizes species-integrated energetic ion flux data from the PS-Out and EP-Out detectors. Consequently, references to the PS and EP unit in this work pertain exclusively to the PS-Out and EP-Out detectors. The fluxes measured by the STEPS detectors are converted from the spacecraft frame to the Geocentric Solar Ecliptic (GSE) coordinate system. The three (PS, IM and EP) out of four remaining detectors measure primarily in the X-Y plane of Geocentric Solar Ecliptic (GSE) coordinate system whereas the NP detector is directed toward the +ve Z direction. STEPS measures the flux of suprathermal/energetic ions in the energy range spanning from 20 keV/n to 6 MeV/n. We use the linear (in a log-log curve of flux vs. energy) part of the spectra to derive the spectral slopes from the fluxes. This leads to selection of ion fluxes with lower cut-off of 0.3 MeV/n for PS and EP detectors and 0.12 MeV/n for IM and NP detectors.  Further, fluxes above 1.3 MeV/n for PS and EP or above 1.2 MeV/n for IM and NP are also not used. These upper cut-offs in the energies for the detectors are chosen to ensure that the present investigation pertains to quiet time. For elemental abundance of suprathermal ions, data of $^3\mathrm{He}$, $^4\mathrm{He}$, C, Fe and O from the Ultra-Low Energy Isotope Spectrometer (ULEIS; \citealt{Stone_et_al_1998}) on board the Advanced Composition Explorer (ACE) spacecraft are used. Since the objective of this work is to investigate quiet time, the criterion for selection of quiet time is described in the subsequent section.

\section{Methodology} \label{sec:Methodology}

This section is divided into two subsections. The first subsection deals with the methodology based on which the quiet time are chosen and the second subsection contains the method on how spectral slopes are calculated.

\subsection{Selection of quiet time}  %%\label{subsec:Selection of quiet time} 

By quiet time, we refer to the conditions in the IP medium when there is no enhancement in the suprathermal ion fluxes beyond a threshold level due to the effects of transient events like SIR/CIRs, ICMEs etc. Quiet times are selected following a heuristic statistical methodology suggested earlier by \citet{Dayeh_et_al_2017}. In this method, as a first step, hourly averaged directional differential flux is calculated based on the measurements by AL1-ASPEX-STEPS-PS, IM, EP, and NP detectors in the energy ranges of $0.36 – 1.32$ MeV/n, $0.13 – 1.22$ MeV/n, $0.38 – 1.32$ MeV/n, and $0.12 – 1.23$ MeV/n, respectively. Subsequently, the fluxes from each direction are sorted in an ascending order. These hourly-averaged and sorted fluxes are re-binned with bin size of 24 hours and the mean value as well as variance are calculated. The mean versus variance values of sorted fluxes for each detector of AL1-ASPEX-STEPS are plotted in Figure 1. For this analysis, data from January 2024 to November 2024 were taken from AL1-ASPEX-STEPS to identify the quiet periods based on the mean–variance technique.

%% The "ht!" tells LaTeX to put the figure "here" first, at the "top" next
%% and to override the normal way of calculating a float position.
%% The asterisk after "figure" tells the compiler to span multiple columns
%% if a two column style is selected.
\begin{figure*}[ht!]
\plotone{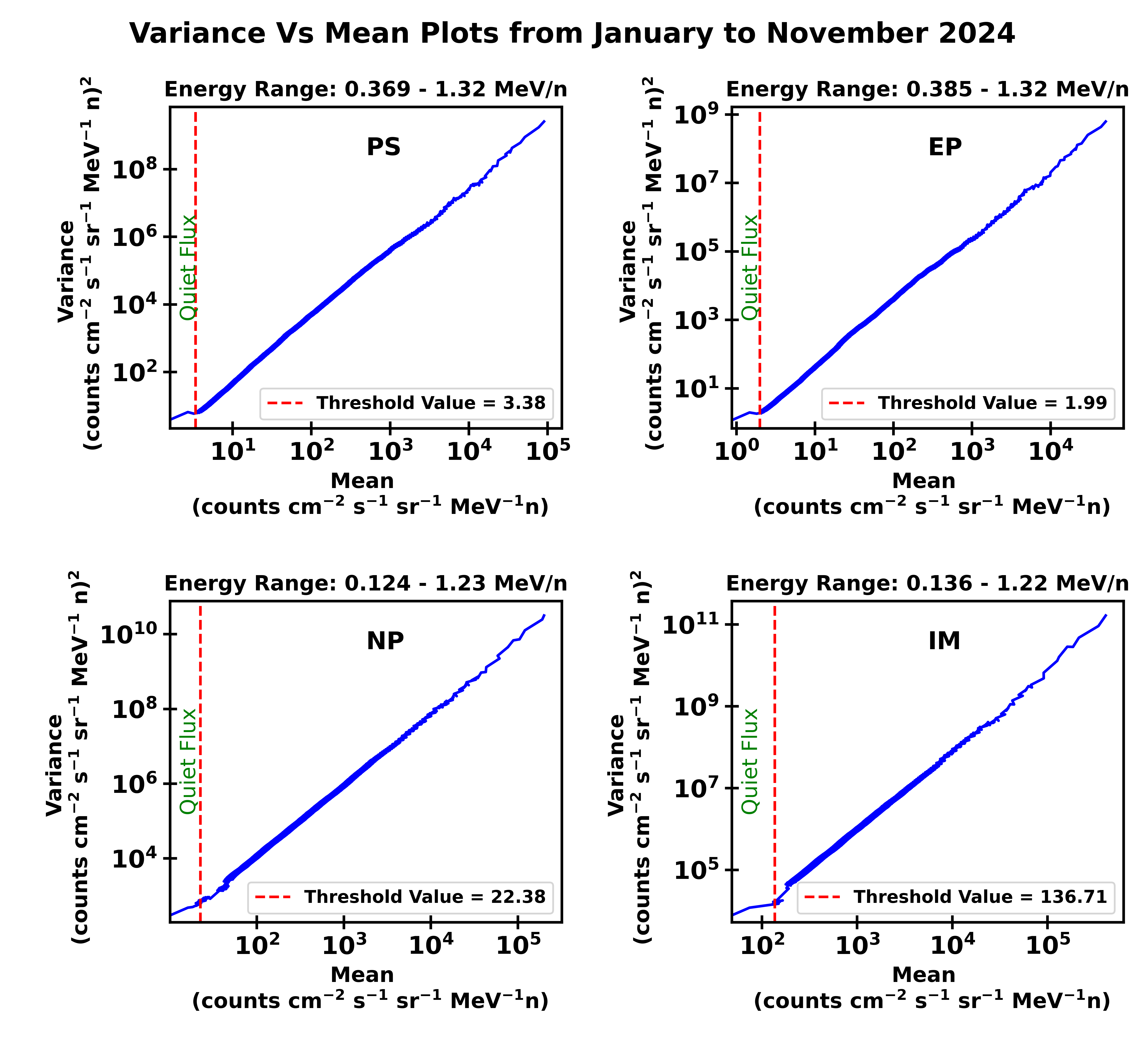}
\caption{Mean versus Variance plots of the sorted (in ascending order) directional differential flux for Parker Spiral (PS), Earth Pointing (EP), North Pointing (NP), and Intermediate (IM) detectors of AL1-ASPEX-STEPS during the observational period from January to November 2024.  The vertical red dashed line in each panel represents the threshold value of the flux below which the fluxes are considered as the quiet time flux and beyond which the fluxes are considered to be associated with transient events. The threshold values are also mentioned. The corresponding energy range is written at the top of each panel.
\label{fig:general}}
\end{figure*}

The vertical dashed red line in each subplot corresponds to the threshold value of the mean flux beyond which the variance in flux increases monotonically. During transient events (CMEs, CIRs/SIRs), particle fluxes exhibit significant enhancements, leading to higher mean values. At the same time, these events are characterized by large fluctuations in fluxes, which results in a monotonically increasing variance with respect to mean values. In contrast, during quiet solar wind periods, the mean flux remains nearly constant with minimal variations, and consequently, no systematic increase in variance with respect to the mean is observed. We have verified that many of these transient enhancements are associated with gradual and impulsive SEP (GSEP/ISEP) events, with flux enhancements ranging from $10^{2}$ to $10^{5}$ for the PS and EP detectors, and from $10^{3}$ to $10^{6}$ for the IM and NP detectors. The characterization of some of these events is complex and will require detailed analyses, which are beyond the scope of the present work; such events will be investigated in future studies. Hence, it is apparent that the mean flux values beyond the threshold value represent enhanced fluxes during transient events (CMEs, CIRs/SIRs), whereas the mean flux values below the threshold value correspond to the quiet-time flux of suprathermal ions in the interplanetary medium. The quiet time identified for each of the detectors for the month of March 2024 is shown in Figure 2.

%% The "ht!" tells LaTeX to put the figure "here" first, at the "top" next
%% and to override the normal way of calculating a float position.
%% The asterisk after "figure" tells the compiler to span multiple columns
%% if a two column style is selected.
\begin{figure*}[ht!]
\plotone{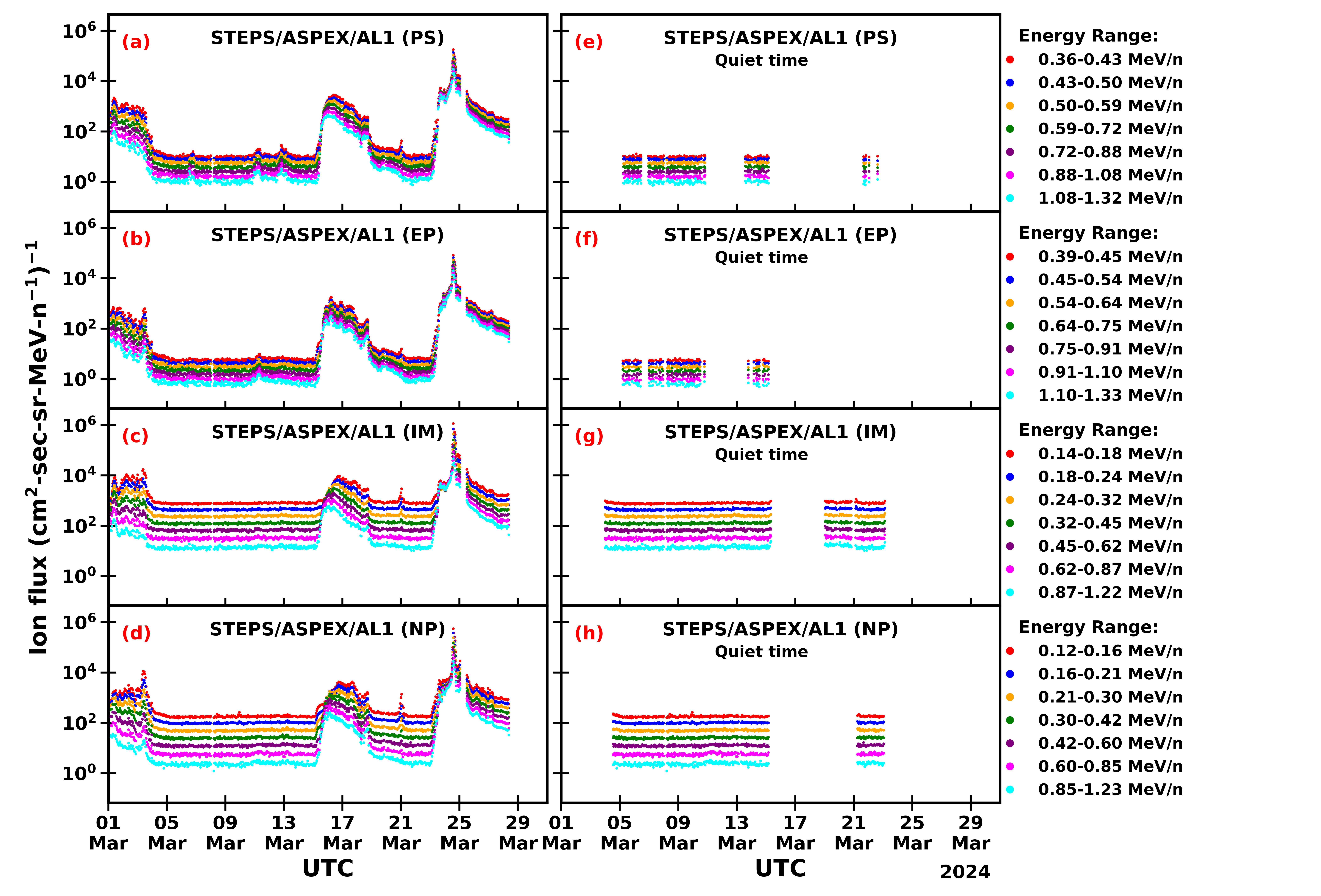}
\caption{Panels a, b, c and d present the temporal variations of differential directional fluxes of suprathermal ions measured by PS, EP, IM and NP detectors of AL1-ASPEX-STEPS, respectively. Panels e, f, g, and h present the corresponding quiet time directional differential flux of suprathermal ions in PS, EP, IM and NP detectors, respectively. The energy channels for the panels are mentioned at the right side of the figure.
\label{fig:general}}
\end{figure*}

In Figure 2, the time-flux profile for March 2024 (panels a, b, c, d) is juxtaposed with the quiet time fluxes identified using the mean vs. variance technique described earlier (panels e, f, g, h).  This methodology is applied for the period from January 2024 to November 2024 except for the month of April 2024 when the satellite underwent many rotations and thruster firings. Spectral indices of suprathermal ions have been calculated for all the quiet time obtained using the above-mentioned methodology and this is discussed in the next subsection. 

\subsection{Estimation of spectral index} %%\label{subsec:Estimation of spectral index}

\begin{figure*}[ht!]
\plotone{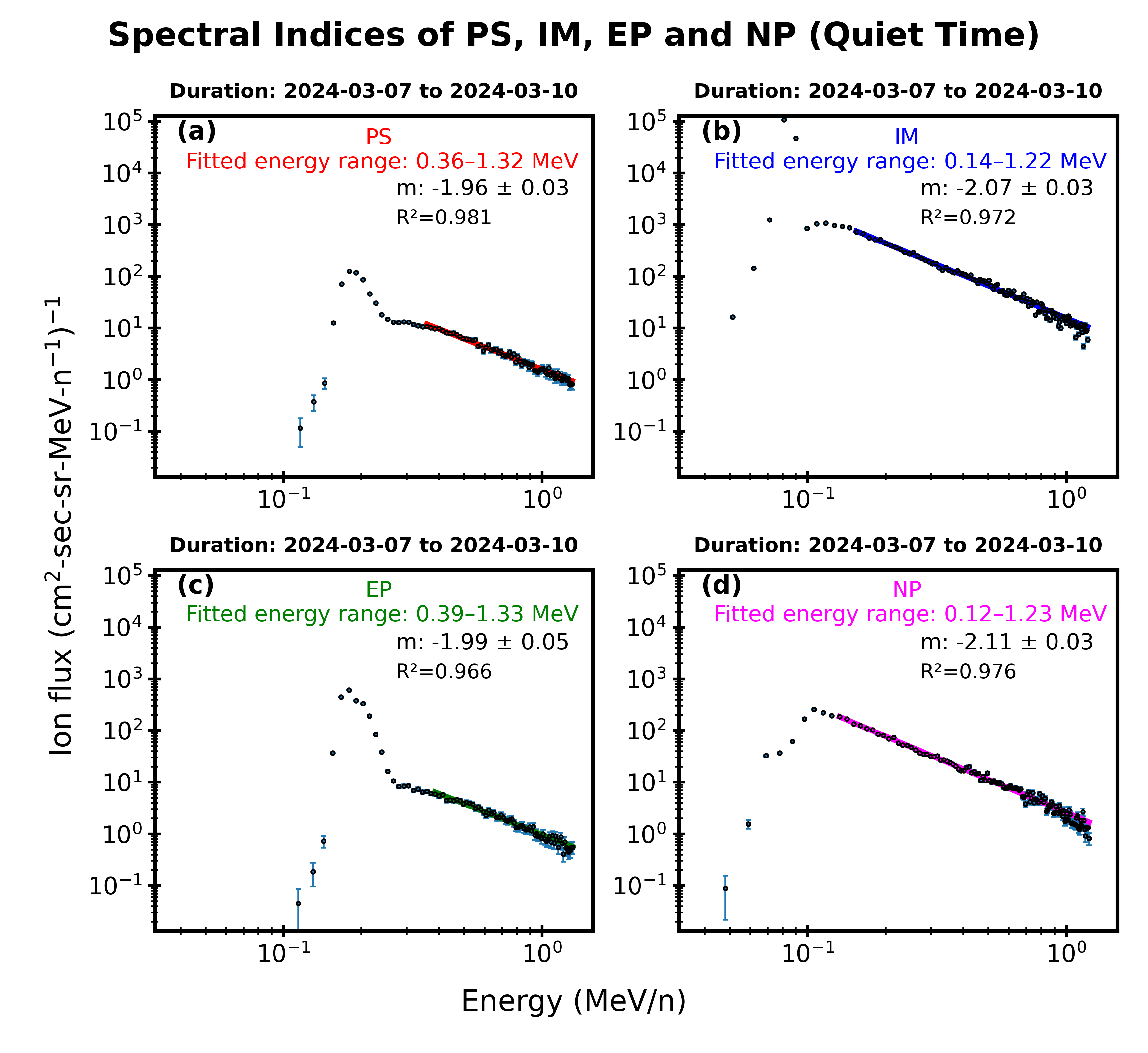}
\caption{Plots of quiet time differential directional flux versus energy with spectral index and     coefficient of determination ($R^2$) values mentioned in each subplot for the PS (a), IM (b), EP (c) and NP (d) directions of AL1-ASPEX-STEPS. Spectral indices are calculated by fitting only the linear part in this log-log plots. The range of energy for the fitted linear part is mentioned in each subplot for PS, IM, EP and NP in red, blue, green and magenta colors respectively.}
\label{fig:general}
\end{figure*}

In figure 3, we show the typical Ion spectra for PS, IM, EP and NP detectors during 07 March, 2024 to 10 March 2024. As can be seen, the nature of the spectra is different at the lower energies. This is due to low-level discriminator (LLD) threshold, which is set electronically to reduce the instrument background noise. The counts below this LLD threshold are considered to be zero and above this value, there is non-linear behaviour up to a certain energy as seen in Figure 3. Towards the low-energy end of the spectrum, the flux distribution exhibits an initial enhancement just above the threshold, followed by a fall before settling into the real physical trend. This initial rise arises due to the response of the solid-state detectors and electronics employing Schmitt trigger based Lower-Level Discriminator (LLD) circuit in the instrument. The LLD sets a minimum amplitude threshold on the detector signals to suppress electronic noise and very low-energy background counts, thereby ensuring reliable measurements of the real signals. However, due to the hysteresis in the values of the LLD cut off, spurious counts from incomplete charge collection and residual electronic noise still pass, leading to an apparent rise in flux values. Different values of low energy enhancements as seen in different detectors are further dependent on the detector types such as the PS and EP detectors, both of which have a dead layer of about $0.8~\mu\mathrm{m}$, display similar low-energy distortions, while the IM and NP detectors, with thinner dead layers of about $0.2~\mu\mathrm{m}$, show a comparable response pattern distinct from the PS and EP detectors. This effect has been further verified during in-orbit payload verification phase, by setting different LLD values for the detectors. At present, the final set LLD values are different for different detectors based on the electronics noise covering the possible lowest energy range. 
Thus, the apparent enhancement and fall observed at the lowest energies are instrument induced rather than of physical origin. Beyond this region as the energy increases, the signal becomes clean and influence of such artefacts diminishes, which subsequently follows the expected power-law trend. We fit \( J = A E^{-m} \) function to the linear portions of the spectra to calculate the spectral indices. To determine the linear portion of the spectra, we evaluated the coefficient of determination ($R^2$) by systematically varying the lower bound of the fitting interval, starting from the point where the spectra visually exhibit a linear trend. The lower energy bound corresponding to the highest ($R^2$) value, obtained after fitting a power-law function, was adopted as the starting energy of the linear portion. The upper energy limits of linear portion of the spectra for the PS, IM, EP, and NP detectors were taken to be 1.32 MeV/n, 1.22 MeV/n, 1.33 MeV/n and 1.23 MeV/n respectively. Thus, the linear portion of the spectra (in log-log space) for PS, IM, EP and NP detectors are found to be within the energy ranges of 0.36-1.32 MeV/n, 0.14-1.22 MeV/n, 0.39-1.33 MeV/n and 0.12-1.23 MeV/n respectively. The energy bins are calculated within these linear fit regions. Thereafter, spectral indices along with the standard errors (fitting errors) are calculated using these energy bins. In this analysis, it is to be noted that the counting uncertainties are also considered while estimating the spectral indices. The counting uncertainties in the flux values are shown as vertical error bars. Since the vertical error bars are very small, these are not clearly visible in log-log scale. 

\section{Results} \label{sec:Results}

In this section, the spectral indices are calculated for all the four ASPEX-STEPS detectors (PS, IM, EP, and NP) under three different scenarios.   As the quiet time are not always identical for all the four detectors, in section 4.1, we identify common quiet time and derive the spectral indices for the four detectors. In section 4.2, we focus on individual detectors and consider all the quiet time for that detector. Further, we also consider a period when the Aditya-L1 spacecraft was rotated for some time and therefore, the four STEPS detectors also rotated from their nominal positions during this interval. This scenario (section 4.3) is brought in to verify whether the spectral indices derived in the first two scenarios remain consistent in the third scenario. The spectral indices derived for these three scenarios are presented in the ensuing subsections.

\subsection{Common quiet time for PS, IM, EP and NP detectors} %%\label{Common quiet time for PS, IM, EP and NP detectors}

We present in Figure 4 the spectral fits and the corresponding indices calculated for a few common quiet time recorded by the PS, IM, EP and NP detectors. The duration of these intervals ranges from a day to a few days as can be seen from the top of each of the plots. Each row in Figure 4 represents a particular detector but for 5 different intervals. The values of spectral indices (with standard fitting errors) for the common quiet-time interval across all detectors are found to be nearly similar, within the range of $-1.99 \pm 0.06$ to $-2.15 \pm 0.02$ (average value $-2.07 \pm 0.04$), suggesting nearly identical spectral indices in all directions.

\begin{figure}[h!]
    \centering
    \includegraphics[width=0.92\textwidth]{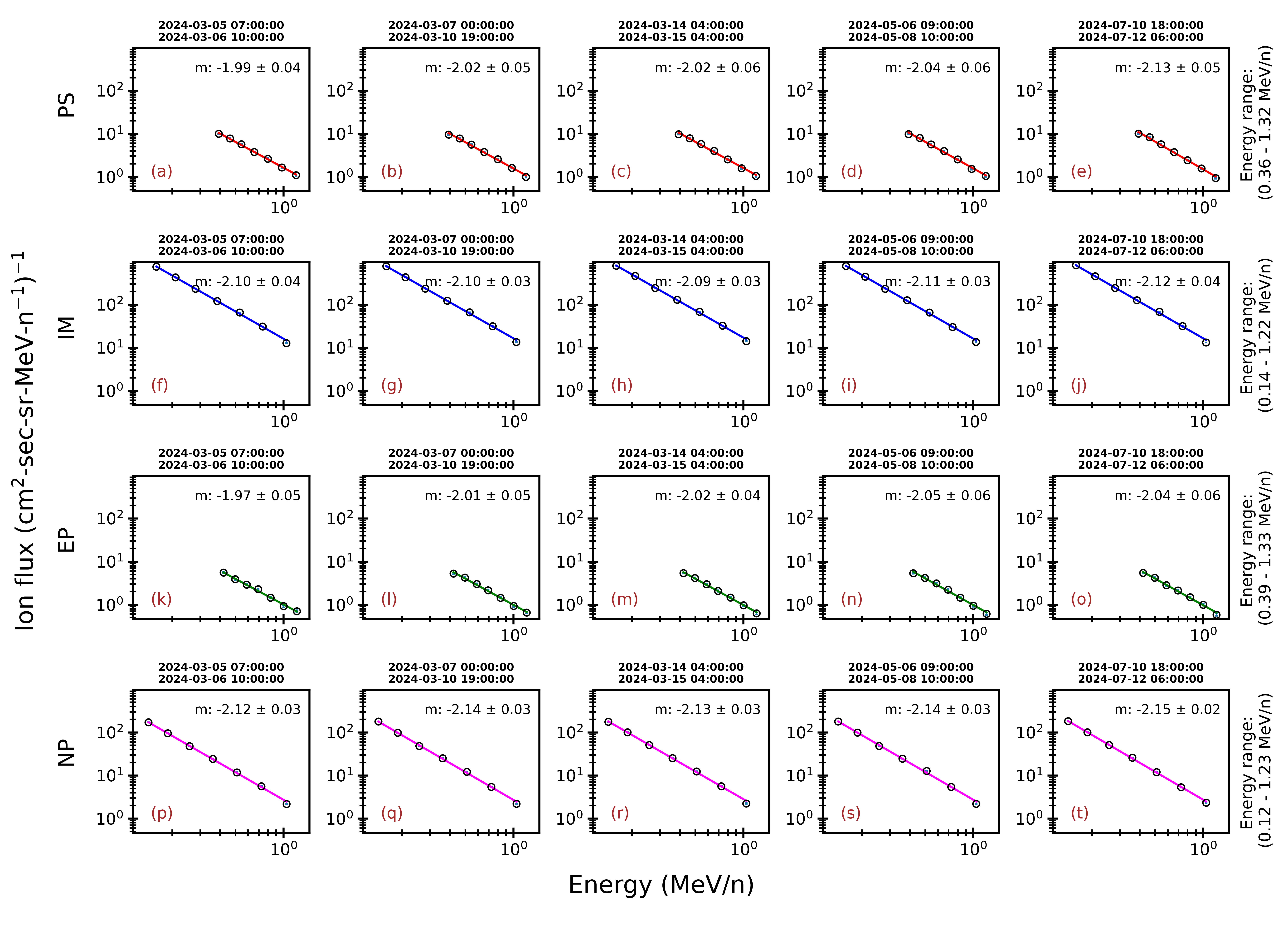}
    \caption{Plots of quiet-time differential directional flux versus energy with spectral index values shown in each subplot for the PS (a–e), IM (f–j), EP (k–o), and NP (p–t) detectors of AL1-ASPEX-STEPS.}
    \label{fig:general}
\end{figure}

\subsection{Quiet time for individual detectors (PS, IM, EP and NP)} %%\label{Quiet time for individual detectors (PS, IM, EP and NP)}

In this section, we present (Figures 5-8) the spectral fits as well as indices calculated for the quiet time selected for each individual detector. Similar to Figure 4, the quiet time are listed at the top of each plots. The spectral indices calculated for quiet time of each detector also show similar values within the range of $-1.84 \pm 0.09$ to $-2.16 \pm 0.02$ (average value $-2.0 \pm 0.06$). Therefore, regardless of the intervals and detectors, the spectral index values are found to be similar and close to $ \sim -2.0$.

\begin{figure*}[ht!]
    \centering
    \includegraphics[width=0.68\textwidth]{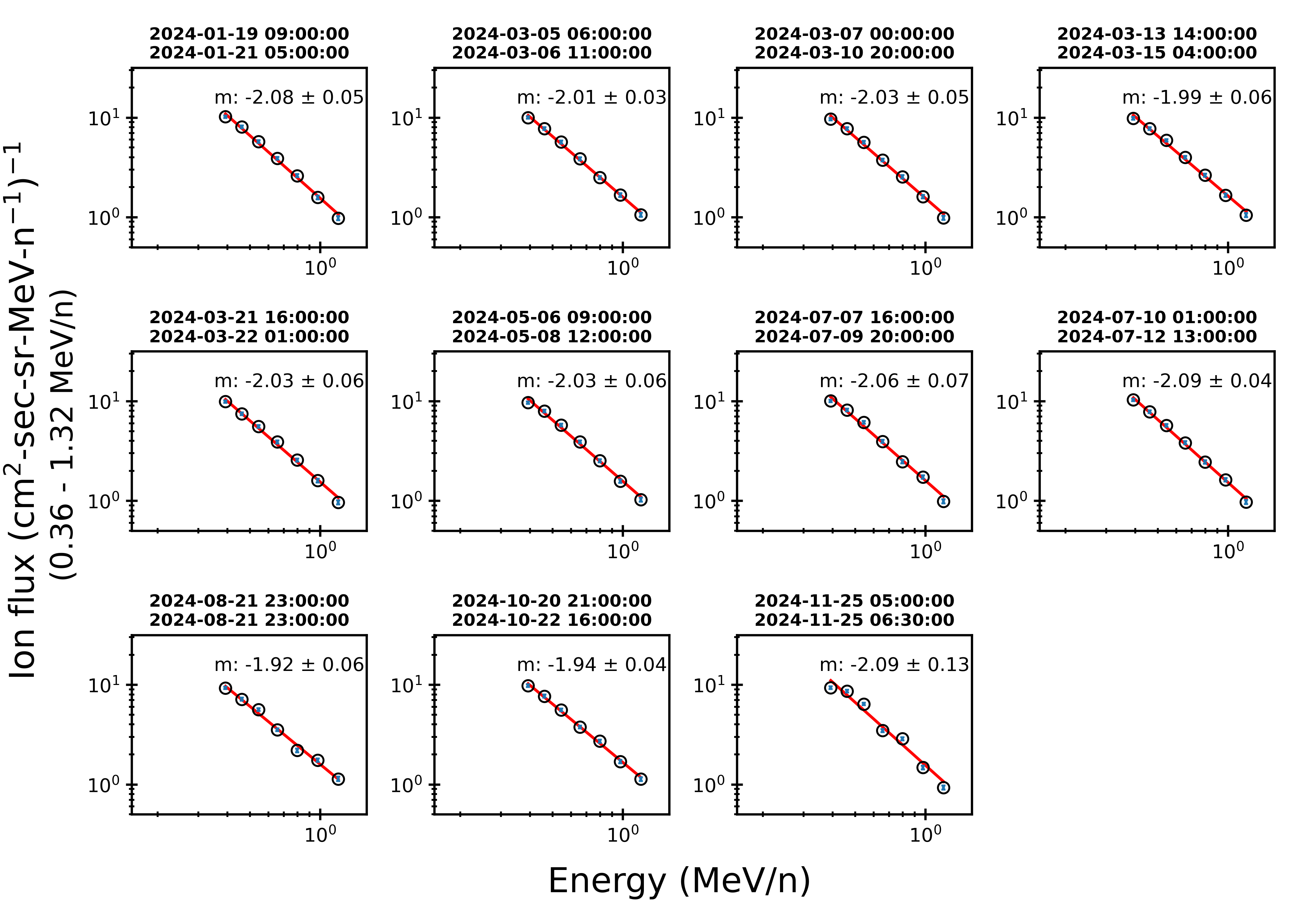}
    \caption{Plots of quiet time differential directional flux versus energy with spectral indices value mentioned in each subplot for the Parker Spiral (PS) detector of AL1-ASPEX-STEPS.
    \label{fig:general}}
\end{figure*}

\begin{figure*}[ht!]
\includegraphics[width=\textwidth]{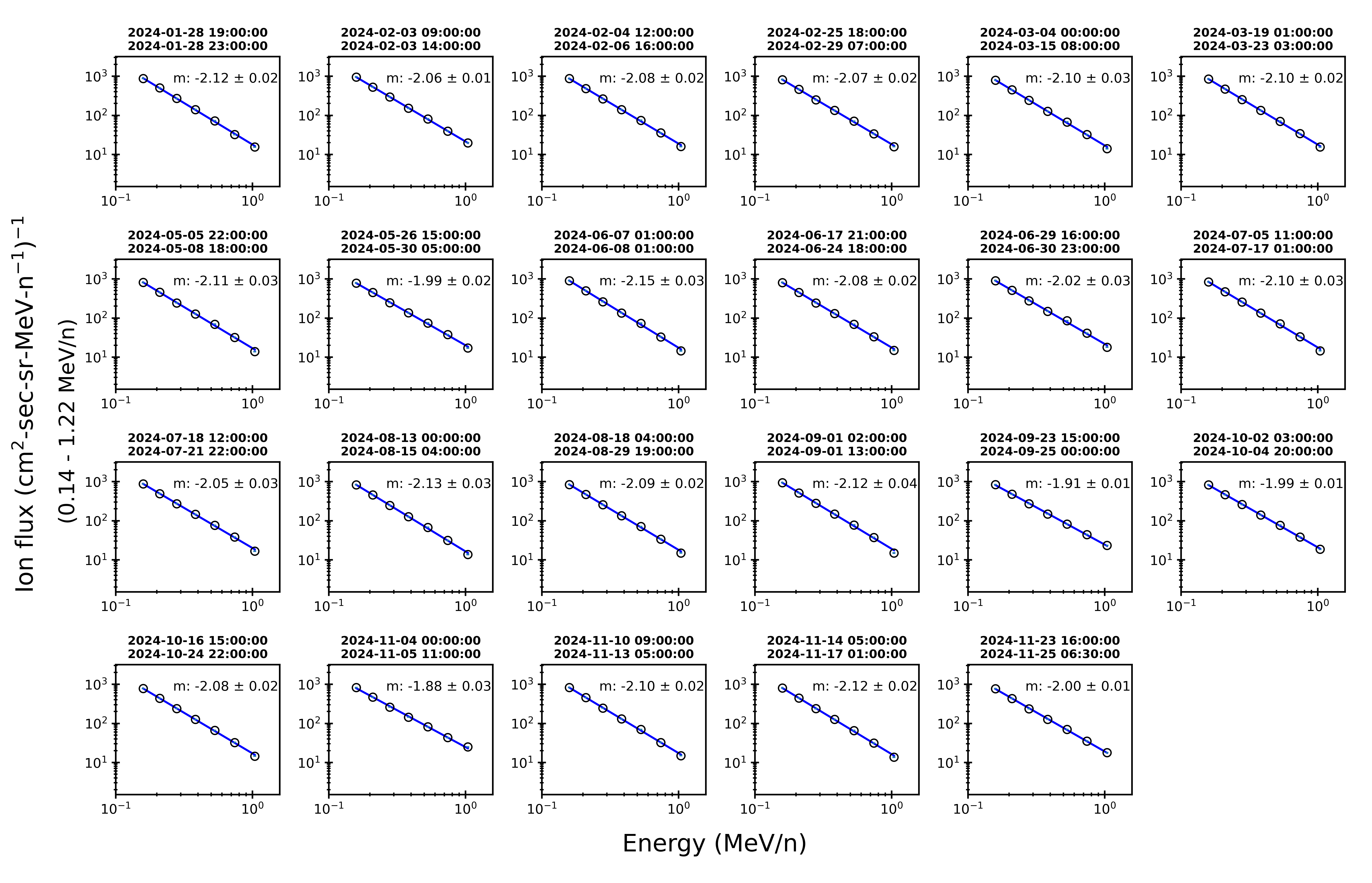}
\caption{Plots of quiet time differential directional flux versus energy with spectral indices value mentioned in each subplot for Intermediate (IM) detector of AL1-ASPEX-STEPS.
\label{fig:general}}
\end{figure*}

\begin{figure*}[ht!]
    \centering
    \includegraphics[width=0.68\textwidth]{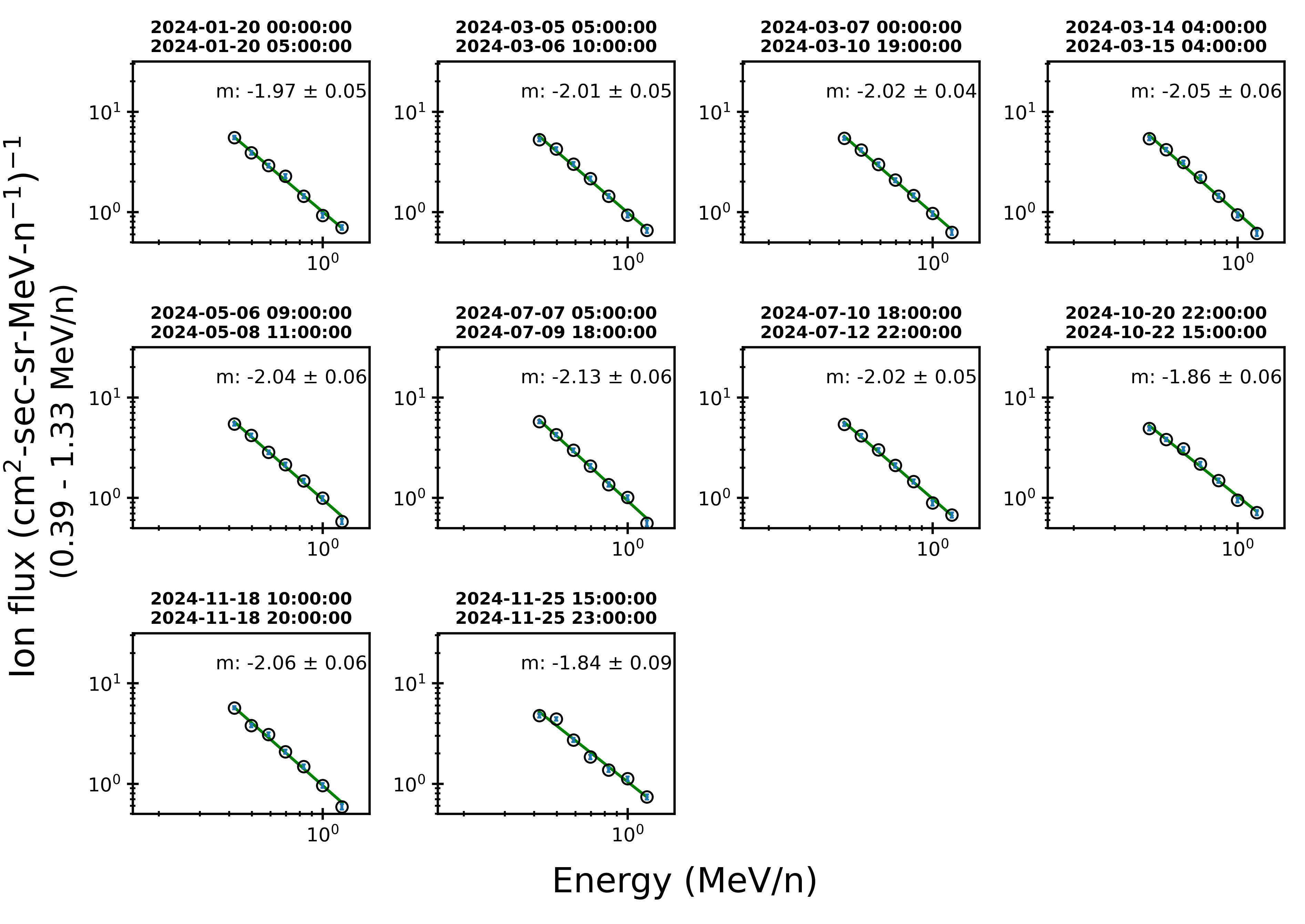}
    \caption{Plots of quiet time differential directional flux versus energy with spectral indices value mentioned in each subplot for Earth pointing (EP) detector of AL1-ASPEX-STEPS.
    \label{fig:general}}
\end{figure*}

\begin{figure*}[ht!]
   \centering
   \includegraphics[width=0.85\textwidth]{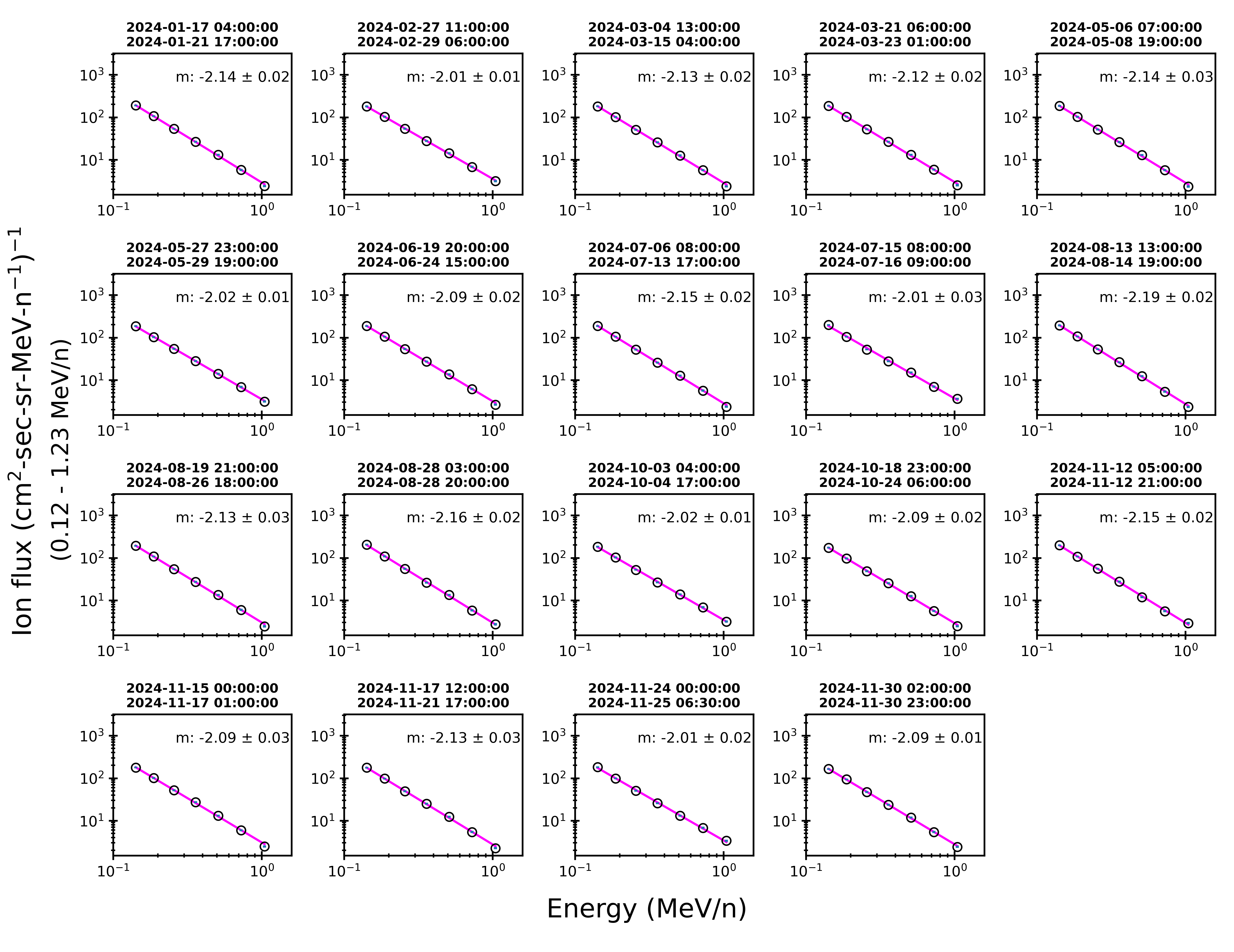}
   \caption{Plots of quiet time differential directional flux versus energy with spectral indices value mentioned in each subplot for North pointing (NP) detector of AL1-ASPEX-STEPS.
   \label{fig:general}}
\end{figure*}

\subsection{Quiet time when AL1 was rotated (during November - 2024)} %%\label{Quiet time when AL1 was rotated (during November - 2024)}

During the quiet interval of 25 November 2024, Aditya L1 (AL1) spacecraft underwent a rotation for a few hours (06:30 UTC to 14:10 UTC). In this rotation period, PS, IM, EP and NP detectors also rotated from their nominal positions. Through Figure 9, we illustrate the orientations of the PS, IM, NP, and EP detectors in their nominal and rotated configurations, represented in the Geocentric Solar Ecliptic (GSE) coordinate system. Panels (a–h) show the nominal configuration, while panels (i–p) display the rotated configuration. In all panels, the X-axis represents the direction cosine corresponding to the $X_{GSE}$ component for each of the four detectors. Panels (a–d) and (i–l) present the direction cosines of the $Y_{GSE}$ component along the Y-axis, whereas panels (e–h) and (m–p) show the direction cosines of the $Z_{GSE}$ component along the Y-axis. Each panel includes a blue and a yellow dot denoting the Earth and the Sun, respectively. The red and green arrows indicate the orientations of the PS, IM, NP, and EP detectors projected in the X–Y and X–Z planes of the GSE coordinate system, respectively. The length of each arrow is proportional to the corresponding direction cosine component, thus accurately reflecting the detector orientation. We estimated the spectral indices for the PS, IM and NP detectors during this quiet interval when the spacecraft was rotated (shown in Figure 10). For the EP detector, the quiet time criteria did not get satisfied and hence spectral indices are not calculated for the EP detector during this period. Interestingly, spectral indices calculated during this interval also show the similar values within the range of $-1.99 \pm 0.04$ to $-2.12 \pm 0.03$ (average value $-2.05 \pm 0.04$) regardless of the detectors and directions. In summary, the Figures 4-10 reveal that spectral indices are almost similar ($\sim -2.0$) regardless of the direction and detector during quiet time.

\begin{figure*}[ht!]
\plotone{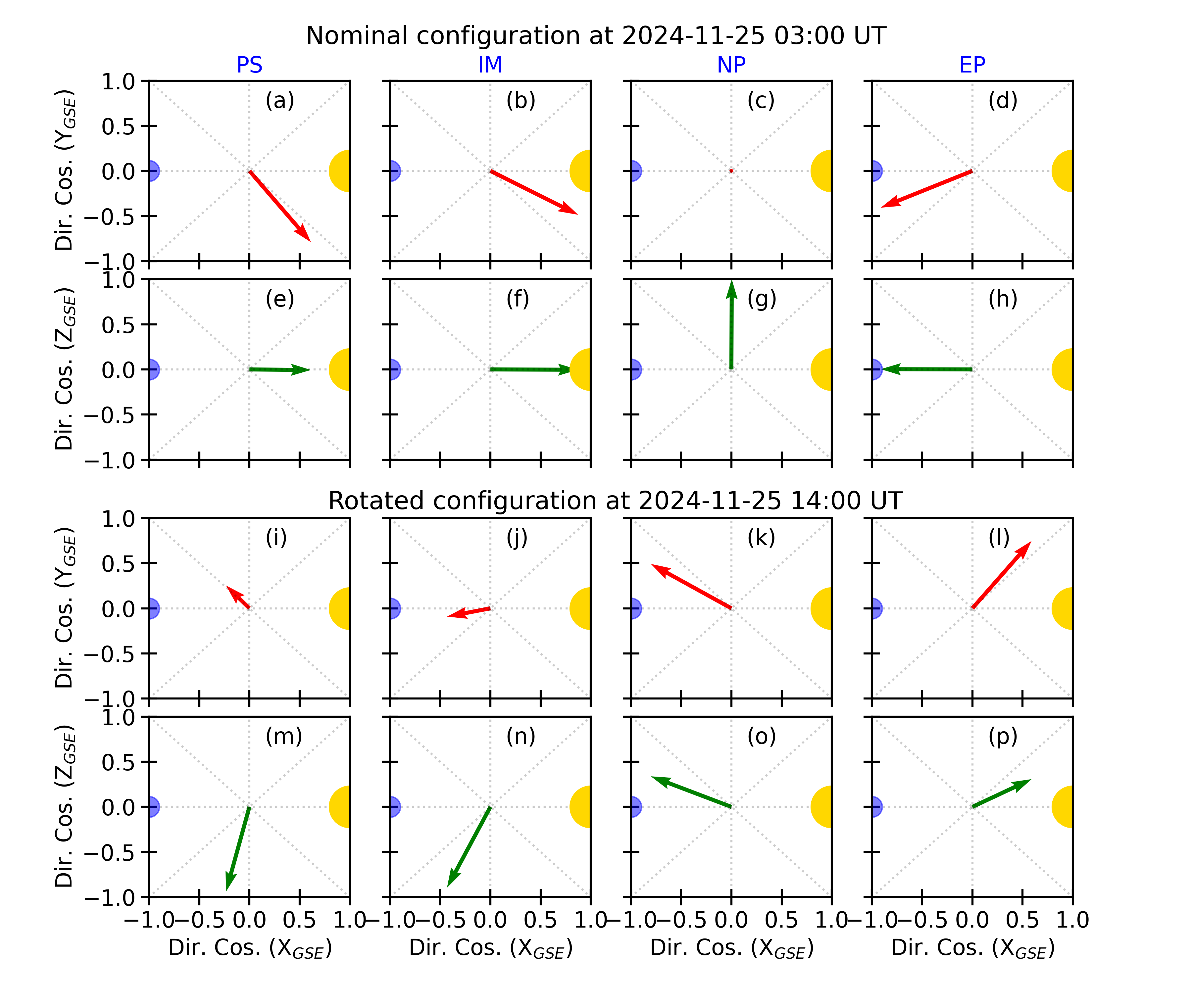}
\caption{Direction cosines of PS, IM, NP and EP detectors in nominal configuration (panel a-h) and in rotated configuration (panel i-p) of AL1-ASPEX-STEPS. In plots (a-p) blue dot and yellow dot represents the Earth and the Sun respectively. Red and green arrows represent the directions of PS, IM, NP and EP detectors in XY and XZ plane of GSE coordinate system respectively. The length of an arrow is decided by the corresponding values of direction cosines.
\label{fig:general}}
\end{figure*}

\begin{figure*}[ht!]
   \centering
   \includegraphics[width=0.75\textwidth]{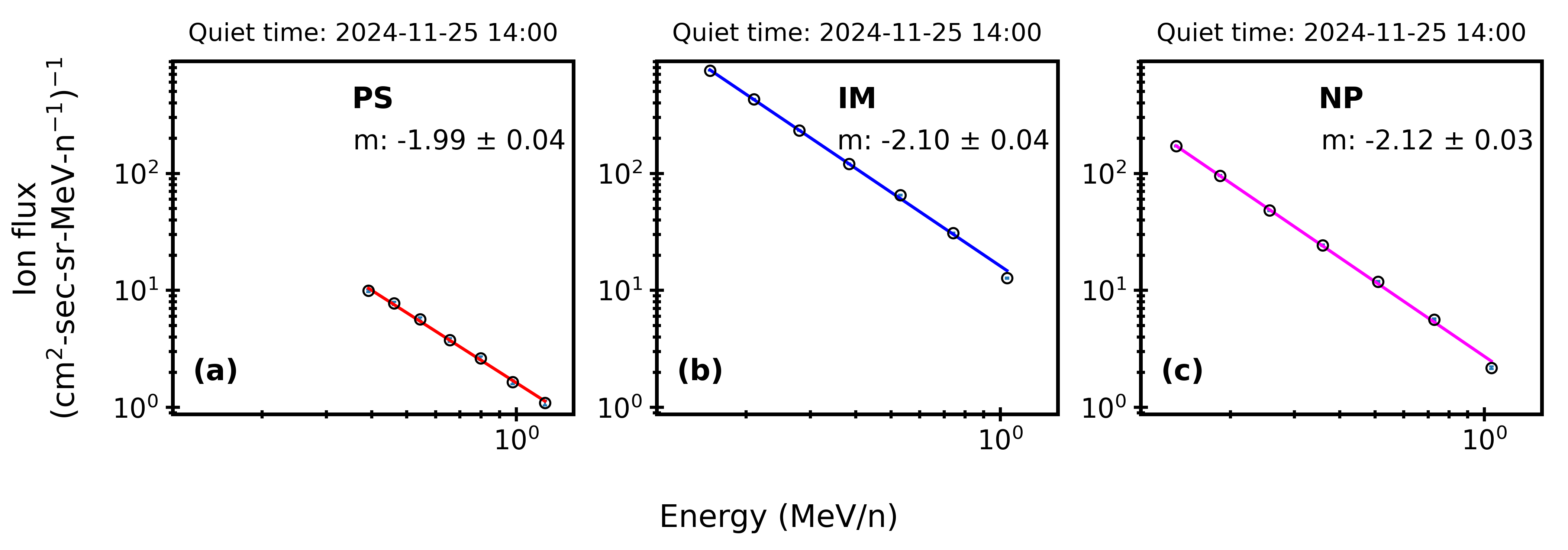}
   \caption{Plots of quiet time differential directional flux versus energy with spectral indices value mentioned in each subplot for PS, IM and NP detectors of AL1-ASPEX-STEPS during rotated period.
   \label{fig:general}}
\end{figure*}

\section{Discussions} \label{sec:Discussions}

In this study, we have calculated spectral indices for quiet time suprathermal ions in IP medium using measurements from four detectors on STEPS instrument of ASPEX onboard Aditya L1 spacecraft during nominal and rotated configurations of the spacecraft. The value of spectral indices obtained from PS, IM, EP and NP detectors are found to be within the range of $-1.84 \pm 0.09$ to $-2.15 \pm 0.02$ (average value $-1.99 \pm 0.06$) that suggests isotropic distribution of quiet time suprathermal ions. A few earlier studies (e.g. \citealt{Dayeh_et_al_2009, Dayeh_et_al_2017, Dalal_et_al_2022}) have shown that the spectral indices vary over a wide range even for quiet periods. These studies were based on the longer intervals of data obtained from instruments onboard spinning spacecraft such as Wind and Advanced Composition Explorer (ACE).  Our analysis is done for shorter intervals of measurements in a directionally resolved manner. In order to assess whether this apparent isotropy reflects a genuine physical property of the ST ions distribution or is influenced by the relative motion between the spacecraft and the solar wind frame, a correction was applied using the Compton-Getting effect \citep{ipavich_1974}. The Compton-Getting effect accounts for the alteration in the observed energy spectra of particles due to the Doppler-like shift arising from the motion of the observer with respect to the solar wind rest frame. This correction enables a more accurate inference of the intrinsic particle distribution by eliminating systematic effects related to the observer's velocity. After considering the Compton-Getting effect, we obtain the corrected or effective spectral index, which better represents the true properties of the particles.

\begin{equation}
m_{\text{eff}} = m' + \frac{(m' + 1)\left( \frac{w}{v} \right) \left[ \cos \theta - \left( \frac{w}{v} \right) \right]}{1 - 2\left( \frac{w}{v} \right)\cos \theta + \left( \frac{w}{v} \right)^2}
\end{equation}

Where,\\
$m_{\text{eff}}$ = Spectral index in the observer’s frame (Spacecraft frame)\\
$m'$ = Spectral index in the solar wind frame\\
$w$ = Relative speed between solar wind and spacecraft\\
$v$ = Speed of the particles/ions\\
$\theta$ = Angle between the solar wind velocity and the direction in which the observer is looking. \\

In this analysis, since the speed of suprathermal ions $\left( v \right)$ observed is much larger than the relative speed $\left( w \right)$ between solar wind and spacecraft,  the term $\left( \frac{w}{v} \right)^2$ in the equation (1) becomes negligible. That’s why we applied the Compton- Getting effect up to the first order correction of $\left( \frac{w}{v} \right)$ in calculating the spectral indices. Equation (1) up to the first order of $\left( \frac{w}{v} \right)$ is given below.

\begin{equation}
m_{\text{eff}} = m' + \frac{(m' + 1)\left( \frac{w}{v} \right)\cos \theta}{1 - 2\left( \frac{w}{v} \right)\cos \theta}
\end{equation}

Spectral indices obtained by applying the correction due to the Compton- Getting effect do not exhibit significant changes up to the first order and remains almost the same $\sim -2.0$. The reason behind the negligible changes in the spectral indices after applying Compton- Getting effect is that the energies (speeds) of the suprathermal ions are very high in comparison to relative speed $\left( w \right)$ between solar wind and spacecraft which makes the term $\left( \frac{w}{v} \right)$ negligible in equation 2. In our analysis, the angle $\theta$ between the solar wind velocity and the PS, IM, EP and NP direction is $52^\circ$, $30^\circ$, $156^\circ$, and $90^\circ$ respectively (for more details, refer \citealt{Goyal_et_al_2025}). There will be no Compton-Getting correction in the North Pointing (NP) direction as $\theta = 90^\circ$ which makes $\cos \theta$ term zero in equation 2. For all other angles, the Compton-Getting corrections are also minimal. Thus, negligible changes are seen in the values of spectral indices for PS, IM and EP directions after applying the Compton- Getting corrections. This shows that the isotropy is a genuine physical property of the suprathermal ions distribution and this is not influenced by the relative motion between the spacecraft and the solar wind frame. This investigation validates the assumptions of isotropic distribution of quiet time suprathermal ions made by \citet{Fisk_and_Gloeckler_2008} in solving Parker transport equation (PTE) for acceleration of suprathermal ions in the solar wind.

Another important insight that comes out of this study is that the spectral indices remain nearly identical $\left( \sim -2 \right)$ for quiet time if we consider a time scale of a few days. This is in contrast to previous studies that reported significant variability in spectral indices when large intervals were considered. For instance, \citet{Dayeh_et_al_2009, Dayeh_et_al_2017} calculated spectral indices using a one-year average for eighteen years (1998–2015) and observed a wide range of values of spectral indices for different elements. Similarly, \citet{Dalal_et_al_2022} analyzed spectral indices using a two-year average across different phases of the solar cycle such as solar maximum, declining phase, solar minimum, and rising phase—and also reported substantial variations in the spectral index values for different elements. Such variability arises because the degree of acceleration experienced by suprathermal populations can vary depending on species, since it is dependent on $m/q$ \citep{Dalal_et_al_2022}. Quiet-time suprathermal ion pools may also contain remnants of earlier transient events (e.g., SEPs), leading to variability in spectral indices as the abundance of different species at a given place and time may depend on the type of SEP event and their $m/q$ dependent transport along interplanetary magnetic field lines. Moreover, the relative dominance of transient solar and interplanetary drivers such as flares, ICMEs, and SIRs that supply suprathermal ions in the interplanetary medium changes with solar phase. This can cause shifts in the dominant acceleration mechanism, producing variability in the spectral indices of quiet-time suprathermals when long intervals are considered. An additional factor is the definition of threshold levels: earlier works (e.g., \citet{Dayeh_et_al_2017, Dalal_et_al_2022}) calculated thresholds based on one- or two-year averages, but such values themselves can vary across solar phases and cycles. This significant variability in spectral index values therefore indicates that even during the quiet period, different processes dominate in different phases of solar cycles. \citet{Dayeh_et_al_2017} suggest that while impulsive and shock acceleration events dominate during enhanced solar activity periods, stream/corotating interaction regions (SIR/CIR) and heated solar wind processes dominate during low solar activity periods. These authors also suggest that the large variability of spectral indices above $100$ keV/n indicates the dominant role of such large, transient processes. Earlier studies by \citet{Gosling_1996}, \citet{chen_et_al_2015} and \citet{hajra_and_sunny_2022} have shown that SIR/CIRs are more prevalent during the declining and minimum phases of the solar cycle. \citet{Dalal_et_al_2023} analyzed the spectral indices of suprathermal ions associated with multiple SIR/CIRs and found typical values of spectral indices in the range of $-3$ to $–4$, indicating significantly softer spectra. In contrast, our study, conducted during a solar maximum period, reports spectral indices closer to $-2$ in all four directions. This suggests that during solar minimum, the quiet-time suprathermal ion pool is primarily influenced by ions accelerated by CIRs/SIRs \citep{Gosling_1996}, resulting in softer spectra \citep{Dalal_et_al_2023}, while during solar maximum, the suprathermal ion population during quiet periods is dominated by large, impulsive and transient events such as solar energetic particle events (SEPs) \citep{Dayeh_et_al_2009, Dayeh_et_al_2017}, leading to harder spectral indices. In this context, the time scale of a few days is also important. By focusing on short time averaging, fixing a year within a solar cycle, and using species-integrated fluxes, we minimize the variable factors mentioned above, leading to the nearly identical spectral indices recorded by all four detectors of AL1-ASPEX-STEPS. As 2024 is a high solar activity year and AL1-ASPEX-STEPS will continue to make measurements, it will be interesting to see whether this nearly identical spectral index continues to hold during low solar activity periods as well, and if the magnitude of the index changes. Further, since ASPEX-STEPS does not discriminate between species and primarily measures protons and alpha particles (being the most abundant), the $m/q$ dependence of the acceleration processes is effectively averaged out. This makes it difficult to comment on the individual processes that fix the spectral indices close to $-2$ at shorter time scales based on the present set of measurements. Therefore, in order to investigate the dominant processes influencing the isotropic distribution of quiet-time suprathermal ions for the same periods as those of ASPEX-STEPS measurements, elemental abundance ratios were estimated using data from the Ultra-Low Energy Isotope Spectrometer (ULEIS) onboard the Advanced Composition Explorer (ACE) spacecraft. This is done following the same methodology suggested in the work of \citet{Dayeh_et_al_2009, Dayeh_et_al_2017}.

Positioned at the L1 point, ACE is well-suited for this analysis as it provides species-resolved measurements of suprathermal ions. The elemental abundance ratios of ${}^3\mathrm{He}/{}^4\mathrm{He}$, $\mathrm{Fe}/\mathrm{O}$, and $\mathrm{C}/\mathrm{O}$ for the period from January to November 2024 are shown in Figure 11. Shaded regions in Figure 11 represent the same quiet periods for which spectral indices have been calculated using ASPEX-STEPS data. In Figure 11, black dots present the averaged elemental abundance ratios of ${}^3\mathrm{He}/{}^4\mathrm{He}$, $\mathrm{Fe}/\mathrm{O}$, and $\mathrm{C}/\mathrm{O}$ during quiet periods, with reference values associated with different particle populations—such as impulsive solar energetic particles (ISEPs) \citep{Mason_2007, Reames_2021}, gradual solar energetic particles (GSEPs) \citep{Mason_2007, Reames_2021}, corotating interaction regions (CIRs) \citep{Mason_et_al_2008}, and the solar wind \citep{Gloeckler_and_Geiss_1998}—indicated by horizontal lines. 

\begin{figure*}[ht!]
\plotone{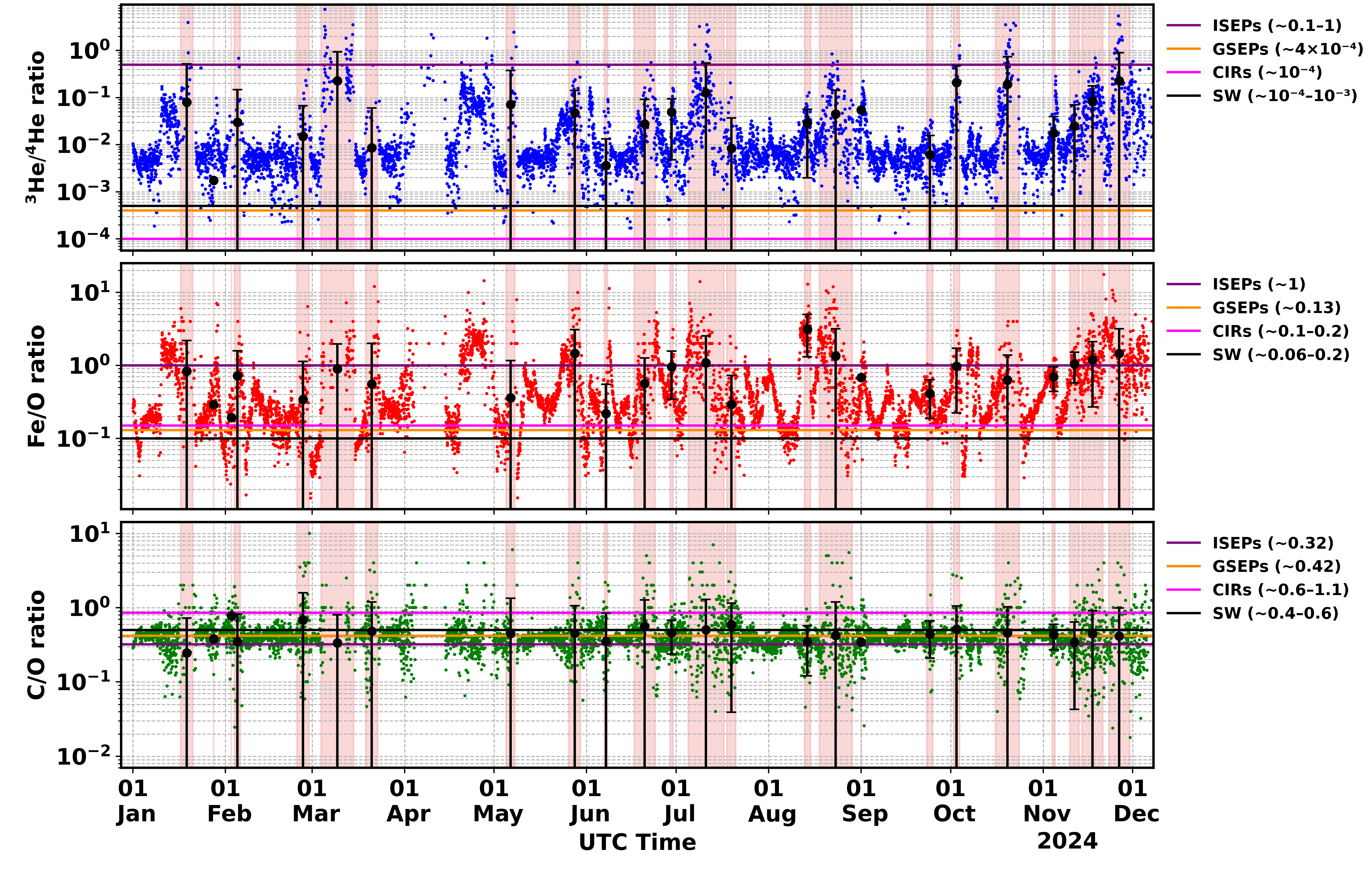}
\caption{Plots of ${}^3\mathrm{He}/{}^4\mathrm{He}$, $\mathrm{Fe}/\mathrm{O}$, and $\mathrm{C}/\mathrm{O}$ ratios as measured by ULEIS/ACE in the energy range spanning from $0.1\,\mathrm{MeV/n}$ to $1.2\,\mathrm{MeV/n}$. The pink shaded regions represent the same quiet periods that are analyzed using the ASPEX-STEPS data. The horizontal lines represent averaged abundances measured in different heliosphere particle populations- CIRs abundance (CIRs; \citealt{Mason_et_al_2008}); ${}^3\mathrm{He}$-rich or impulsive SEP events (impulsive SEPs; \citealt{Mason_2007, Reames_2021}); gradual SEP events (GSEPs) (gradual SEPs; \citealt{Mason_2007, Reames_2021}); Solar Wind (SW) value \citep{Gloeckler_and_Geiss_1998}. Black dots in each subplot represent the average value of abundance ratios during quiet periods with one sigma error bar. 
\label{fig:general}}
\end{figure*}

Figure 11 reveals that the quiet-time suprathermal ion population is enriched in both ${}^3\mathrm{He}$ and $\mathrm{Fe}$. This enrichment is consistent with earlier observations of elevated ${}^3\mathrm{He}$ \citep{Bucik_et_al_2014, Bucik_et_al_2015} and $\mathrm{Fe}$ \citep{Gloeckler_1975} during quiet periods. Based on Figure 11, the values of elemental abundance ratios for quiet periods are listed in Table 1 for reference. Table 1 presents a summary of elemental abundance ratios observed during selected quiet time intervals in 2024. For each interval, the average and standard deviation of three key elemental abundance ratios—${}^3\mathrm{He}/{}^4\mathrm{He}$, $\mathrm{Fe}/\mathrm{O}$, and $\mathrm{C}/\mathrm{O}$—are provided. The start and end of each interval are mentioned with date and time \text{(DD:MM:YYYY HH:MM)} in the first and second column of the table respectively. The ${}^3\mathrm{He}/{}^4\mathrm{He}$ ratio is a sensitive tracer of impulsive solar energetic particle (ISEP) events and during these events, enhanced ${}^3\mathrm{He}$
 is commonly observed \citep{Reames_1999, Mason_et_al_1999, Desai_et_al_2016}. The $\mathrm{Fe}/\mathrm{O}$ ratio serves as an indicator of charge-to-mass-dependent acceleration processes \citep{Reames_1995, Reames_1999, Tylka_et_al_2006}, while the $\mathrm{C}/\mathrm{O}$ ratio provides complementary information about the relative abundance of lighter ions \citep{Reames_1999, Von_Steiger_et_al_2000}.

% In the document body:
\begin{deluxetable*}{cccccccc}
\tablecaption{Quiet time elemental abundance ratio of ${}^3\mathrm{He}/{}^4\mathrm{He}$, Fe/O, and C/O from ULEIS/ACE from 0.1 MeV/n to 1.2 MeV/n with standard deviation values.}
\tablehead{
\colhead{Start date} & \colhead{End date} & 
\colhead{Avg $^3$He/$^4$He} & \colhead{Std $^3$He/$^4$He} & 
\colhead{Avg Fe/O} & \colhead{Std Fe/O} & 
\colhead{Avg C/O} & \colhead{Std C/O}
}
\startdata
17-01-2024 04:00 & 21-01-2024 17:00 & 0.069 & 0.408 & 0.612 & 1.073 & 0.279 & 0.493 \\
28-01-2024 19:00 & 28-01-2024 23:00 & 0.028 & 0.040 & 1.813 & 2.957 & 0.647 & 0.416 \\
03-02-2024 09:00 & 03-02-2024 14:00 & 0.003 & 0.007 & 0.219 & 0.091 & 0.488 & 0.345 \\
04-02-2024 12:00 & 06-02-2024 16:00 & 0.031 & 0.112 & 0.913 & 0.772 & 0.223 & 0.287 \\
25-02-2024 18:00 & 29-02-2024 07:00 & 0.017 & 0.056 & 0.387 & 0.846 & 0.828 & 1.401 \\
04-03-2024 00:00 & 15-03-2024 08:00 & 0.219 & 0.698 & 0.892 & 1.074 & 0.327 & 0.456 \\
19-03-2024 01:00 & 23-03-2024 03:00 & 0.008 & 0.051 & 0.647 & 1.622 & 0.471 & 0.708 \\
05-05-2024 22:00 & 08-05-2024 19:00 & 0.095 & 0.342 & 0.757 & 1.623 & 0.418 & 1.092 \\
26-05-2024 15:00 & 30-05-2024 05:00 & 0.050 & 0.098 & 1.516 & 1.703 & 0.451 & 0.634 \\
07-06-2024 01:00 & 08-06-2024 01:00 & 0.006 & 0.020 & 0.232 & 0.329 & 0.365 & 0.490 \\
17-06-2024 21:00 & 24-06-2024 18:00 & 0.029 & 0.065 & 0.847 & 0.998 & 0.528 & 0.728 \\
29-06-2024 16:00 & 30-06-2024 23:00 & 0.046 & 0.050 & 0.695 & 0.663 & 0.457 & 0.225 \\
05-07-2024 11:00 & 17-07-2024 01:00 & 0.131 & 0.425 & 1.039 & 1.474 & 0.517 & 0.806 \\
18-07-2024 12:00 & 21-07-2024 22:00 & 0.008 & 0.026 & 0.304 & 0.442 & 0.586 & 0.527 \\
13-08-2024 00:00 & 15-08-2024 04:00 & 0.028 & 0.026 & 2.967 & 1.864 & 0.343 & 0.218 \\
18-08-2024 04:00 & 29-08-2024 19:00 & 0.043 & 0.097 & 1.233 & 1.778 & 0.419 & 0.749 \\
01-09-2024 02:00 & 01-09-2024 13:00 & 0.061 & 0.047 & 0.650 & 0.354 & 0.570 & 0.475 \\
23-09-2024 15:00 & 25-09-2024 00:00 & 0.006 & 0.011 & 0.408 & 0.225 & 0.407 & 0.260 \\
02-10-2024 03:00 & 04-10-2024 20:00 & 0.152 & 0.256 & 0.752 & 0.735 & 0.512 & 0.559 \\
16-10-2024 15:00 & 24-10-2024 22:00 & 0.184 & 0.537 & 0.615 & 0.760 & 0.446 & 0.589 \\
04-11-2024 00:00 & 05-11-2024 11:00 & 0.057 & 0.073 & 0.706 & 0.266 & 0.440 & 0.152 \\
10-11-2024 09:00 & 13-11-2024 05:00 & 0.025 & 0.046 & 1.082 & 0.509 & 0.367 & 0.331 \\
14-11-2024 05:00 & 21-11-2024 17:00 & 0.080 & 0.091 & 1.407 & 1.569 & 0.448 & 0.531 \\
23-11-2024 16:00 & 30-11-2024 23:00 & 0.251 & 0.649 & 1.356 & 1.648 & 0.410 & 0.571 \\
\enddata
\end{deluxetable*}

Table 1 reveals characteristic signatures of impulsive solar energetic particle (ISEP) contributions during several intervals that are otherwise considered quiet. Specifically, substantial enhancements in the ${}^3\mathrm{He}/{}^4\mathrm{He}$ ratio are observed during many quiet intervals that are commonly associated with ISEPs. Some of these intervals are 04–15 March, 05–17 July, 02–04 October, 16–24 October, 14–21 November, and 23–30 November 2024. The elevated ${}^3\mathrm{He}/{}^4\mathrm{He}$
 values  during these intervals, reaching or exceeding 0.1, suggest that approximately 25\% of the selected quiet time are influenced by residual impulsive SEPs from previous events. In addition, the $\mathrm{Fe}/\mathrm{O}$ ratio exhibits considerable variability throughout the year. Notably, elevated $\mathrm{Fe}/\mathrm{O}$ values greater than 2, as seen on 28 January and during 13–15 August 2024, provides strong indication on the presence of ISEP remnants in the suprathermal ion pool. It is seen that approximately 45\% of the quiet periods display $\mathrm{Fe}/\mathrm{O}$ ratios near unity (Fe/O $\sim 1$), indicating the contribution of ISEP events, which are characterized by enhanced heavy ion abundances. The $\mathrm{C}/\mathrm{O}$ ratio adds further insight into the compositional makeup of the suprathermal ion population. Approximately 20\% of the quiet time show $\mathrm{C}/\mathrm{O}$ values near 0.32, indicative of impulsive SEP contributions, while around 50\% of the intervals display values near 0.42, which are characteristic of gradual SEP populations. Together, these findings underscore the complexity of particle composition even during nominally quiet times and highlight the persistent influence of impulsive events in determining the compositional makeup of the suprathermal ion populations.

\section{Summary} \label{sec:Summary}

In this study we focus on short-term quiet time of suprathermal ions during solar maximum and the analysis leads to the following key findings:

\begin{enumerate}
    \item The distribution of suprathermal ions during quiet time is nearly isotropic, as indicated by the similar spectral index values within the range of $-1.84 \pm 0.09$ to $-2.15 \pm 0.02$ (average value $-1.99 \pm 0.06$) observed in all four directions—Parker spiral (PS), intermediate (IM), Earth pointing (EP), and north pointing (NP).

    \item This near isotropic behavior supports the assumption made by \citet{Fisk_and_Gloeckler_2008}, who considered an isotropic suprathermal ion distribution while solving the Parker transport equation to explain the acceleration of suprathermal ions.

    \item We infer that this near isotropic behavior comes out owing to the consideration of species-integrated fluxes, short time-averaging and fixing a year within a solar cycle, These choices  minimize the variable factors that may contribute to the quiet time variability of the spectral indices.

    \item Elemental abundance ratios indicate that the quiet-time suprathermal ion population has significant contributions from solar energetic particle (SEP) events: approximately 25\% from impulsive SEPs based on the ${}^3\mathrm{He}/{}^4\mathrm{He}$ ratio, about 45\% from impulsive SEPs inferred from the $\mathrm{Fe}/\mathrm{O}$ ratio, and nearly 20\% from impulsive SEPs along with ~50\% from gradual SEPs as suggested by the $\mathrm{C}/\mathrm{O}$ ratio.

    \item The suprathermal ion pool during quiet time are predominantly composed of residual particles from previous impulsive and gradual solar energetic particles (SEPs) events in solar maximum conditions.

\end{enumerate}

%% Please use the acknowledgment and contribution environments. This will 
%% be anonomyized when the "anonymous" style option is used. 
\begin{acknowledgments}
Authors also would like to thank various centres of ISRO for providing technical support and facilities for the test and calibration of ASPEX payload. Thanks are also due to the project, mission, and review teams of ISRO for their support. Authors are also grateful to the principal investigators and members of the ACE-ULEIS team for generating and managing the datasets used in this work. We also sincerely acknowledge Dr. Mihir Desai from the Southwest Research Institute (SwRI) for discussions during the course of this work.
\end{acknowledgments}

\section*{DATA AVAILABILITY STATEMENTS} 

The data used in this study are publicly available. The ASPEX-STEPS data from the Aditya-L1 mission are accessible from two sources: data from January 7 to May 10, 2024, are available via Zenodo (\doi{10.5281/zenodo.15833452}) \citep{chakrabarty_2025_15833452}, and data from May 11, 2024, onwards can be obtained from the ISSDC PRADAN portal (\url{https://pradan.issdc.gov.in/al1/}). The ULEIS data from the ACE spacecraft are available through NASA’s CDAWeb (\url{https://cdaweb.gsfc.nasa.gov/}) under the DOI \doi{10.48322/agmq-ex61}
 \citep{mason_uleis}.

\bibliography{Aakash_Manuscript_08072025}{}
\bibliographystyle{aasjournalv7}

\end{document}